\documentclass[10pt, aps, prb, twocolumn, showpacs, floatfix,superscriptaddress]{revtex4-1}

\usepackage{amsmath, amssymb, mathtools, array, color, bbm}
\usepackage{hyperref}
\hypersetup{colorlinks=true, linkcolor=blue, citecolor=blue}
\usepackage{graphicx}
\usepackage{times}

\newcommand{ \bra }			[1] { \langle{#1}| }
\newcommand{ \ket }				[1] { |{#1}\rangle }

\newcommand{ \ex }				[1] { \langle{#1}\rangle }

\newcommand{ \com }			[1] { \left[ {#1} \right] }

\newcommand{\trm}				{\textrm}

\newcommand{\pd}				{\phantom{\dag}}

\newcommand{ \op }				{ \hat{p} }
\newcommand{ \ox }				{ \hat{x} }

\newcommand{ \tr }				{{\rm{Tr}}}

\newcommand{\nn}				{\nonumber}

\newcommand{\fv}				{\mathcal{V}}
\newcommand{\fw}				{\mathcal{W}}

\newcommand{\ve}				{\varepsilon}

\begin{document}


\title{Transport properties of double quantum dots with electron-phonon coupling}

\author{Stefan~Walter}
\affiliation{Department of Physics, University of Basel, Klingelbergstrasse 82, CH-4056 Basel, Switzerland}

\author{Bj{\"o}rn~Trauzettel}
\affiliation{Institute for Theoretical Physics and Astrophysics, University of W{\"u}rzburg, 97074 W{\"u}rzburg, Germany}

\author{Thomas~L.~Schmidt}
\affiliation{Department of Physics, University of Basel, Klingelbergstrasse 82, CH-4056 Basel, Switzerland}

\date{\today}


\begin{abstract}
We study transport through a double quantum dot system in which each quantum dot is coupled to a phonon mode.
Such a system can be realized, e.g., using a suspended carbon nanotube. We find that the interplay between strong
electron-phonon coupling and inter-dot tunneling can lead to a negative differential conductance at bias voltages
exceeding the phonon frequency. Various transport properties are discussed, and we explain the physics of the
occurrence of negative differential conductance in this system.
\end{abstract}

\maketitle

\section{Introduction}
\label{sec:intro}

Over the past decades, it has become clear that quantum dot systems are ideally suited for a detailed study of electronic
transport phenomena in mesoscopic physics. Notable transport features through single quantum dots include the Coulomb
blockade effect,\cite{Beenakker1991,Averin1991,Kouwenhoven1997, Kouwenhoven2001a} the Kondo effect,\cite{Ng1988, Meir1993,Kouwenhoven2001b}
and the spin blockade effect.\cite{Weinmann1995} Double quantum dots\cite{Wiel2003} are a natural extension. They consist
of two quantum dots connected either in parallel or in series. One of the most interesting effects found in double dots with
strong electronic interactions is a negative differential conductance if the tunnel couplings to both dots are different.\cite{Fransson2004}

The small size of a quantum dot gives rise to a quantization of its energy levels. As a consequence, transport through
quantum dots at finite bias voltages usually occurs via one or several localized electronic levels in the bias window, and
current and noise measurements can be used as experimental probes of this level structure. The Coulomb repulsion on
the dot also has a strong impact on its transport properties because it limits the number of electrons occupying the dot.
This Coulomb blockade phenomenon has been observed in many experiments on different length scales.

The electronic level structure of a quantum dot depends sensitively on its shape. Therefore, vibrational modes of the dot
give rise to interactions between electrons and phonons. The effect of electron-phonon interactions on transport properties
in quantum dot systems have been studied theoretically~\cite{Wingreen1988, Brandes1999, Boese2001, Braig2002, Mitra2003, Brandes2003, Koch2005, Koch2006, Zazunov2006, Egger2008,schmidt09,haupt09,avriller09,maier11,Santamore2013}
and have been observed in numerous experiments on different systems. Electron transport in molecular wire junctions~\cite{Nitzan2003}
can be studied using STM techniques or mechanically controlled break junctions. Single atoms or molecules connected to
two contacts can be prepared and measured as quantum dots. Experiments have been performed, for instance, on
H$_2$ molecules~\cite{Smit2002} and on other more complicated molecules.~\cite{Zhitenev2002,Qiu2004}
Similar effects at other energy scales were observed in experiments on suspended carbon nanotubes~\cite{LeRoy2004,Sazonova2004,Sapmaz2006,Steele2009,Lassagne2009,Leturcq2009} or
in experiments on buckyballs.\cite{Park2000,Pasupathy2005} Even larger systems, e.g., quantum shuttles,~\cite{Gorelik1998}
also fall under the same paradigm. Such nanoelectromechanical systems\cite{Poot2012} make it possible to study the
influence of phonons on transport through the device in a very controllable way. Recently, it has been demonstrated
that it is possible to tailor the interaction between localized electronic degrees of freedom and the mechanical degree of
freedom of a suspended carbon nanotube in a very controlled way.~\cite{Benyamini2013}

In this article, we study transport through a double quantum dot system influenced by the presence of phonons on each dot.
Naively, one expects the current through the double dot system to increase with the applied bias voltage. However, as
we show below, a negative differential conductance can arise for sufficiently strong electron-phonon coupling, i.e., the current
can \emph{decrease} when the bias voltage is increased. Moreover, this negative differential conductance occurs even if the
system is symmetric.

The article is organized as follows. In Sec.~\ref{sec:keyresults}, we propose the model and discuss a possible realization of it.
We furthermore summarize our key results. We formally introduce the Hamiltonian of the underlying model in Sec.~\ref{sec:model}.
In Sec.~\ref{sec:mastereq}, we use a Born-Markov master equation approach to determine the rate equations which can be
used to calculate the current and differential conductance. We present and discuss the results of the current and differential
conductance in Sec.~\ref{sec:currentANDdIdV}. Finally, we summarize in Sec.~\ref{sec:summary}.

\section{Model and key results}
\label{sec:keyresults}

\begin{figure}[t]
	\centering
	\includegraphics[width=0.99\columnwidth]{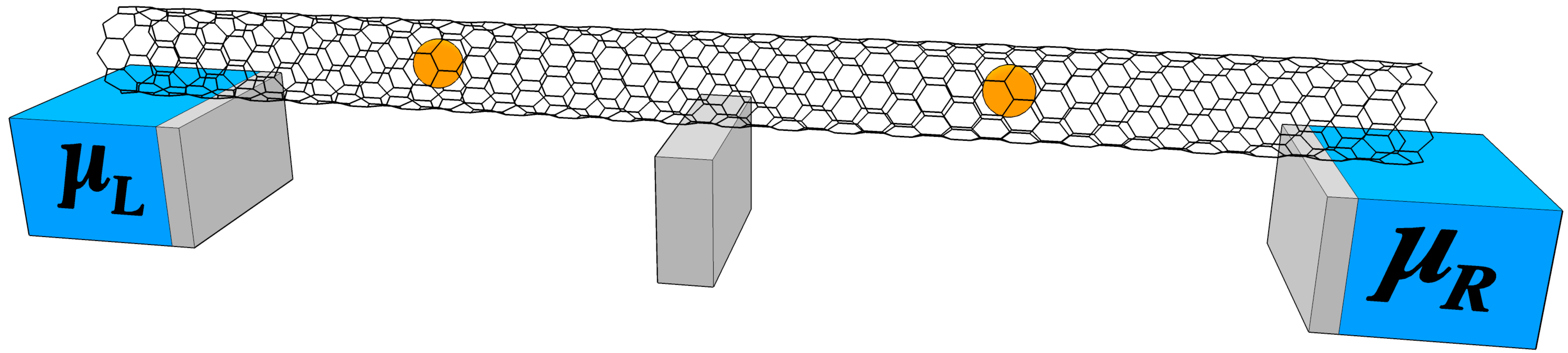}
	\caption{\label{fig:setup} (Color online) A carbon nanotube is suspended between two metallic leads (blue) held at chemical
	potentials $\mu_{L}$ and $\mu_{R}$. The additional support in the center separates the nanotube into two regions; each
	free to oscillate. Tunnel barriers are depicted in gray. The orange spheres	denote quantum dots that form on each suspended
	part of the nanotube.}
\end{figure}

We investigate transport through a double quantum dot setup, in which the energy of each electronic level depends linearly
on the displacement of one phonon mode. Such a system can be realized, e.g., using carbon nanotube (CNT) setups, where
the central part of the CNT is supported, whereas the two lateral parts are suspended, see Fig.~\ref{fig:setup}. The suspended
sections of the CNT serve as quantum dots \cite{Postma2001} with large charging energies, and are free to oscillate. Using a gate voltage, the central part is tuned to an insulating regime, so
transport can only occur if an electron from the left section of the CNT tunnels into the right section. CNTs are especially
favorable for this kind of setup because of (i) their high $Q$-factors and stiffness,~\cite{Huttel2009,Huttel2010} (ii) high
vibrational frequencies in the range of $4-11$ GHz,~\cite{Chaste2011} and (iii) large electron-phonon coupling.\cite{Leturcq2009}
Note, however, that the model we consider is fairly generic, and we expect that it can be realized also using alternative
molecular quantum dot or nanoelectromechanical systems.

The large charging energy and the weak coupling to the metallic contacts allow us to use a rate equation approach, and to take
into account only sequential tunneling processes. As we show below, this is the regime in which a negative differential conductance
in the double dot setup can be observed. We find that for fixed inter-dot tunneling and at bias voltages on the order of the phonon
frequency, the current is suppressed when increasing the electron-phonon coupling. Furthermore, we find that for large electron-phonon
coupling and relatively weak inter-dot coupling, the current decreases when increasing the bias voltage, leading to a negative
differential conductance. This negative differential conductance disappears when the inter-dot coupling is increased. We conclude
that there is an interesting interplay between electron-phonon coupling  and the inter-dot coupling which in certain cases leads to
a negative differential conductance.

\section{Hamiltonian}
\label{sec:model}
\begin{figure}[t]
	\centering
	\includegraphics[width=0.95\columnwidth]{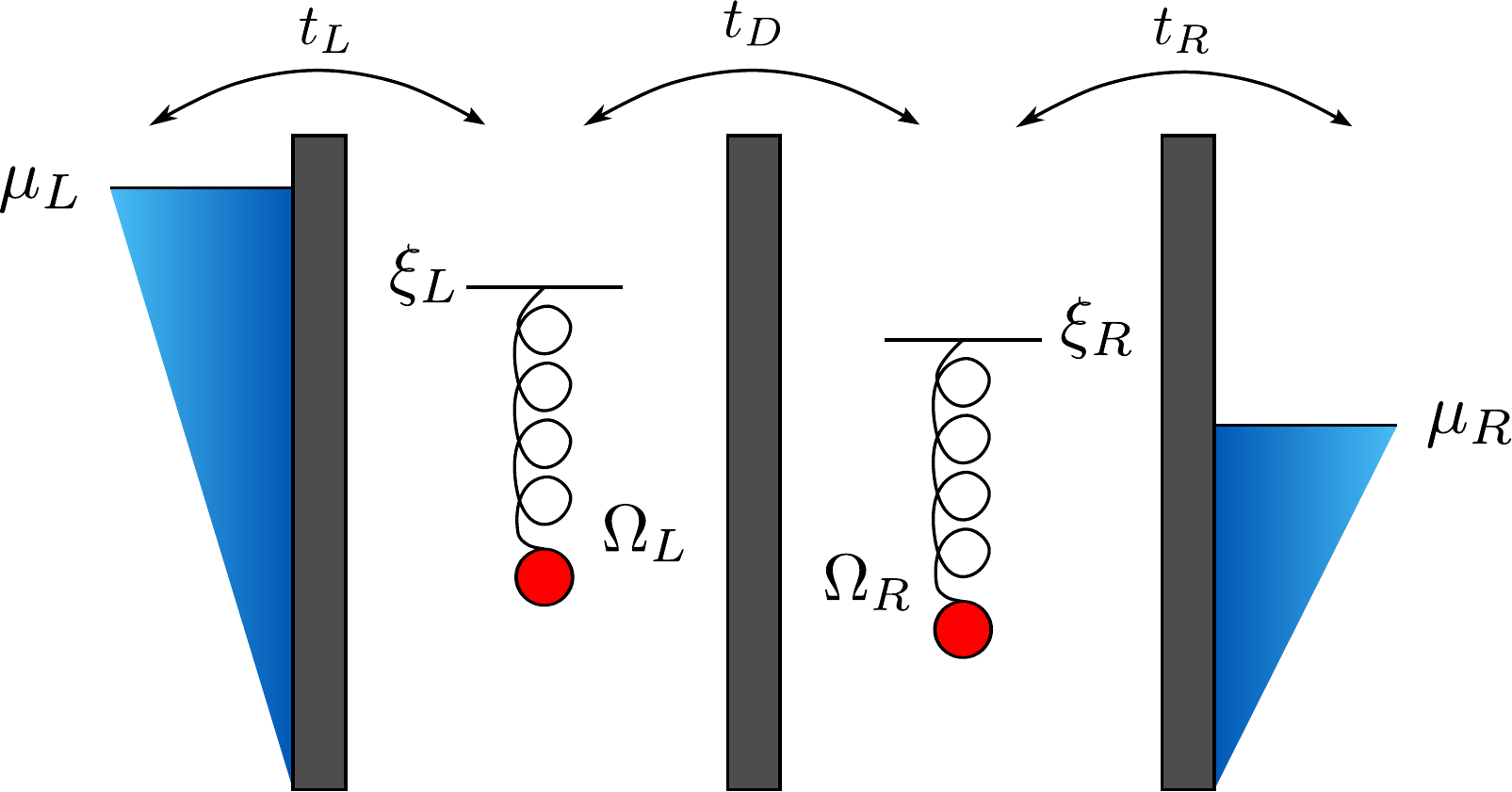}
	\caption{\label{fig:DDph} (Color online)
	Schematic picture of the setup described by the Hamiltonian (\ref{eq:H}). Two quantum dots with onsite energies
	$\xi_{L}$ and $\xi_{R}$ are coupled to each other with a tunneling amplitude $t_{D}$, and to electron reservoirs
	on the left and right with tunneling amplitudes $t_{L}$ and $t_{R}$, respectively. The electron reservoirs are
	normal-metal leads with chemical potentials $\mu_{L}$ and $\mu_{R}$, respectively. A phonon mode with
	frequency $\Omega_{L}$ ($\Omega_{R}$) is coupled to the left (right) dot.}
\end{figure}

%
Figure~\ref{fig:DDph} shows a schematic diagram of the setup we consider in the following: each of the two dots contains
a single electronic level in the bias window, which is coupled to one phonon mode. Two normal-metal leads, held at chemical
potentials $\mu_L$ and $\mu_R$ (i.e. the bias voltage is $V=\mu_{L}-\mu_{R}$), are attached to the double quantum dot to drive a current through the system. The total
Hamiltonian describing this model is given by $(\alpha = \{L,R\})$
\begin{align}\label{eq:H}
	H &= \sum_{\alpha} \left[ H_{\trm{lead}}^{(\alpha)} + H_{\trm{dot}}^{(\alpha)} + H_{\trm{osc}}^{(\alpha)} + H_{\trm{osc-dot}}^{(\alpha)} \right] \nn \\
& + H_{\trm{dd}} + H_{\trm{tun}} \, ,
\end{align}
where the different parts are
\begin{align}
	H_{\trm{lead}}^{(\alpha)} 		&= \sum_{k} \ve_{k}^{\pd} \psi_{\alpha k}^{\dag} \psi_{\alpha k}^{\pd} \, , \nn \\
	H_{\trm{dot}}^{(\alpha)} 		&= \xi_{\alpha}^{\pd} d_{\alpha}^{\dag} d_{\alpha}^{\pd} \, , \nn \\
	H_{\trm{osc}}^{(\alpha)} 		&= \frac{\op_{\alpha}^{2}}{2m_{\alpha}} + \frac{1}{2} m_{\alpha} \Omega_{\alpha}^{2} \ox_{\alpha}^{2} \, , \nn \\
	H_{\trm{osc-dot}}^{(\alpha)} 	&= \lambda_{\alpha}^{\pd} \ox_{\alpha}^{\pd} d_{\alpha}^{\dag} d_{\alpha}^{\pd} \, , \nn \\
	H_{\trm{dd}} 				&= t_{D}^{\pd} d_{L}^{\dag} d_{R}^{\pd} + t_{D}^{\pd} d_{R}^{\dag} d_{L}^{\pd} \, , \nn \\
	H_{\trm{tun}} 				&= \sum_{\alpha,k} t_{\alpha}^{\pd} \psi_{\alpha k}^{\dag} d_{\alpha}^{\pd} + \trm{H.c.} \nn \,
\end{align}
Here, $H_{\trm{lead}}^{(\alpha)}$ describes the normal-metal leads using electron creation and annihilation operators,
$\psi_{\alpha k}^{\dag}$ and $\psi_{\alpha k}^{\pd}$, respectively, for electrons with wave vector $k$ in lead $\alpha$.
The dot Hamiltonian $H_{\trm{dot}}^{(\alpha)}$ describes a single electronic orbital at energy $\xi_{\alpha}$, where
$d_{\alpha}^{\dag}$ $(d_{\alpha}^{\pd})$ creates (annihilates) an electron on dot $\alpha$. The phonons which couple
to the dots are described by the harmonic oscillator Hamiltonian $H_{\trm{osc}}^{(\alpha)}$. The electron-phonon
coupling is given by the Hamiltonian $H_{\trm{osc-dot}}^{(\alpha)}$, where $\lambda_{\alpha}$ denotes the coupling
strength of the phonon mode to the occupation number of dot $\alpha$. The inter-dot coupling is given by $H_{\trm{dd}}$
with tunneling amplitude $t_{D}$. Finally, $H_{\trm{tun}}$ couples each dot to its adjacent normal-metal lead with an
energy-independent tunneling amplitude $t_{\alpha}$.

We assume spin-independent transport and large intra- and inter-dot Coulomb repulsion, such that the double dot
works as a single electron transistor, i.e., only one spinless electron can occupy the double dot system at any given
time. Therefore, the corresponding Hilbert space of the electronic double dot system is spanned by the three states
\begin{align}\label{eq:basis}
	\ket{0,0} &\equiv \ket{0} \, , \nn \\
	\ket{1,0} &\equiv \ket{L} \, , \nn \\
	\ket{0,1} &\equiv \ket{R} \, ,
\end{align}
where $\ket{n_{L},n_{R}}$ denotes a state with $n_{L}$ $(n_{R})$ electrons on the left (right) dot.

The electron-phonon coupling $\lambda_{L,R}$ can be strong, e.g., in an experimental realization employing CNTs.
In order to treat it exactly, we use a polaron (Lang-Firsov) transformation which eliminates the electron-phonon
coupling term in Eq.~(\ref{eq:H}).\cite{Mahan2000} Using the unitary transformation
\begin{align}
	S = \sum_{\alpha} e^{-i \Lambda_{\alpha} \op_{\alpha} n_{\alpha}} \nn \, ,
\end{align}
where $\Lambda_{\alpha} = \lambda_{\alpha}/m_{\alpha}\Omega_{\alpha}^{2}$ and $n_{\alpha} = d_{\alpha}^{\dag} d_{\alpha}^{\pd}$,
the transformed Hamiltonian $\tilde{H}$ reads
\begin{align}
	\tilde{H} &= S H S^{\dag} \nn \\
&= \sum_{\alpha} H_{\trm{lead}}^{(\alpha)} + \tilde{H}_{\trm{dot}}^{(\alpha)} + H_{\trm{osc}}^{(\alpha)} + \tilde{H}_{\trm{dd}} + \tilde{H}_{\trm{tun}} \nn \, ,
\end{align}
where
\begin{align}
	\tilde{H}_{\trm{dot}}^{(\alpha)} 	&= \tilde{\xi}_{\alpha}^{\pd} d_{\alpha}^{\dag} d_{\alpha}^{\pd} \, , \nn \\
	\tilde{H}_{\trm{dd}} 			&= t_{D}^{\pd} d_{L}^{\dag} X_{L}^{\dag} X_{R}^{\pd} d_{R}^{\pd} + t_{D}^{\pd} d_{R}^{\dag} X_{R}^{\dag} X_{L}^{\pd} d_{L}^{\pd} \, , \nn \\
	\tilde{H}_{\trm{tun}} 			&= \sum_{\alpha,k} t_{\alpha}^{\pd} \psi_{\alpha k}^{\dag} d_{\alpha}^{\pd} X_{\alpha}^{\pd} + \trm{H.c.} \nn \, .
\end{align}
As a consequence of the electron-phonon coupling, the level energies are renormalized, $\tilde{\xi}_{\alpha} = \xi_{\alpha} - \Lambda_{\alpha} \lambda_{\alpha} / 2$,
and the polaron operator $X_{\alpha} = e^{i \op_{\alpha} \Lambda_{\alpha}}$ emerges in the electron tunneling Hamiltonian.
The complicated structure of the polaron operator makes an exact solution impossible. Therefore, we shall use a perturbative
approach in the dot-lead tunnel amplitudes $t_{L,R}$ and the inter-dot tunnel amplitude $t_D$.

\section{Born-Markov master equation}
\label{sec:mastereq}

To calculate transport properties of the double dot system for arbitrary electron-phonon coupling, we employ a Born-Markov
master equation approach. We separate the full Hilbert space into system and bath degrees of freedom, where the system
contains the double dot, whereas the lead electrons as well as the phonons form the bath. The Markov approximation consists
in assuming that the bath is in thermal equilibrium at all times. The full density matrix can therefore be approximated as
$\rho_{\trm{tot}}(t) \approx \rho_{\trm{dots}}(t) \otimes \rho_{\trm{ph}} \otimes \rho_{\trm{leads}}$. Moreover, we treat the
tunneling to second order (Born approximation). This also implies that we neglect backaction effects by tunneling on the electrons in the leads and on the phonons. Tracing out the bath degrees of freedom, we arrive at a master equation for
the double dot density matrix (we set $\hbar=1$),
\begin{widetext}
\begin{align}\label{eqn:bmme1}
	\frac{d}{dt} \rho_{\trm{dots}}(t) = -i \tr_{\trm{ph}} \com{ \sum_{\alpha} \tilde{H}_{\trm{dot}}^{(\alpha)} + \tilde{H}_{\trm{dd}}, \rho_{\trm{dots}}(t) }
						- \int_{0}^{\infty} dt'  \tr_{\trm{ph}} \tr_{\trm{leads}} \left\{ \left[ \tilde{H}_{\rm{tun}}, \left[ \tilde{H}_{\rm{tun}}(-t'), \rho_{\trm{dots}}(t) \otimes \rho_{\trm{ph}} \otimes \rho_{\trm{leads}} \right] \right] \right\} \, .
\end{align}
\end{widetext}
We included the inter-dot tunneling term $\tilde{H}_{\trm{dd}}$ in the system part, i.e., it appears in the term describing the
coherent time evolution in Eq.~(\ref{eqn:bmme1}). However, we excluded the effect of $\tilde{H}_{\trm{dd}}$ on the
time-evolution of $\tilde{H}_{\rm{tun}}(-t')$,~\cite{Goan2001} i.e., the time evolution of $\tilde{H}_{\rm{tun}}(-t')$ is in the
interaction picture with respect to the unperturbed Hamiltonian $H_{0} = \sum_{\alpha} H_{\trm{lead}}^{(\alpha)} + \tilde{H}_{\trm{dot}}^{(\alpha)} + H_{\trm{osc}}^{(\alpha)}$.
The latter approximation is justified in the limit $t_{D} \ll  t_{\alpha} \ll \max(eV,k_{B} T_{el})$. Therefore, we also treat
$t_{D}$ as a small perturbation.

\subsection{Rate equations and current through the system}

In the weak tunneling limit that we consider, the tunneling rate from the leads to the dots and vice versa is much smaller
than the phonon energy $\Omega_{\alpha}$. In the stationary case, the occupation probabilities $p_{0}$, $p_L$ and $p_R$
of the three basis states (\ref{eq:basis}) satisfy the following rate equations,
\begin{align}
	0 &= -(W_{0 L} + W_{0 R}) p_{0} +  W_{L 0} p_{L} + W_{R 0} p_{R} \, , \label{eqn:bmme2a} \\
	0 &= W_{0 L} p_{0} - (W_{L 0} p_{L} + W_{L R}) p_{L} + W_{R L} p_{R} \, , \label{eqn:bmme2b} \\
	0 &= W_{0 R} p_{0} + W_{L R} p_{L} - (W_{R 0} + W_{R L}) p_{R} \, . \label{eqn:bmme2c}
\end{align}
where $W_{\alpha  \beta}$ denotes the rate for tunneling from state $\alpha$ to $\beta$ ($\alpha,\beta \in \{0,L,R\}$).
Using Eqs.~(\ref{eqn:bmme2a})-(\ref{eqn:bmme2c}) and the normalization condition $p_{0}+p_{L}+p_{R}=1$, we can
solve for the occupation probabilities $p_{0}$, $p_{L}$, $p_{R}$, and calculate the stationary current
\begin{align}
	I = -e \big[ p_{0} W_{0 R} - p_{R} W_{R 0} \big] \, . \nn
\end{align}
Because of current conservation, it is enough to consider the current from the right dot to the right lead.
The transition rates $W_{\alpha  \beta}$ are obtained from the master equation Eq.~(\ref{eqn:bmme1}).

\subsection{Equation of motion for the density matrix}

We obtain the rates and the current from the matrix elements $\bra{\alpha} \rho_{\trm{dots}}(t) \ket{\beta} = \rho_{\alpha \beta}(t)$
of Eq.~(\ref{eqn:bmme1}). The differential equations for matrix elements are
\begin{align}
	\dot{\rho}_{00} = &- [ W_{0 L} + W_{0 R} ] \rho_{00} + W_{L 0} \rho_{LL} + W_{R 0} \rho_{RR} \, , \label{eqn:rateeqsD1} \\
	\dot{\rho}_{LL} = & - i t_{D} [ M_{\trm{LR}} \rho_{RL} - M_{\trm{RL}} \rho_{LR} ] - [W_{0 R} + W_{L 0} ] \rho_{LL} \nn \\
					&+ W_{0 L} \rho_{00} \, , \label{eqn:rateeqsD2} \\
	\dot{\rho}_{RR} = & - i t_{D} [ M_{\trm{RL}} \rho_{LR} - M_{\trm{LR}}\rho_{RL} ] - [W_{0 L} + W_{R 0} ] \rho_{RR} \nn \\
					&+ W_{0 R} \rho_{00} \, , \label{eqn:rateeqsD3} \\
	\dot{\rho}_{LR}  = &- i t_{D} M_{\trm{LR}} \Big[ \rho_{RR} - \rho_{LL} \Big] - i \Big[ \tilde{\xi}_{L} - \tilde{\xi}_{R} \Big] \rho_{LR} \nn \\
					&- \fw \rho_{LR}/2 \, , \label{eqn:rateeqsD4} \\
	\dot{\rho}_{RL}  = &i t_{D} M_{\trm{RL}} \Big[ \rho_{RR} - \rho_{LL} \Big] + i \Big[ \tilde{\xi}_{L} - \tilde{\xi}_{R} \Big] \rho_{RL}  \nn \\
					&-\fw \rho_{RL}/2 \, , \label{eqn:rateeqsD5}
\end{align}
with $\fw = \left[ W_{0 R} + W_{R 0} + W_{0 L} + W_{L 0} \right]$. The tunneling rates are given by
\begin{align}
	W_{0 \alpha} &= \int_{-\infty}^{\infty} \frac{d\omega}{2\pi} \, \Gamma_{\alpha} f_{\alpha} (\tilde{\xi}_{\alpha} + \omega) F_{\alpha}^{<}(\omega) \, , \nn \\
	W_{\alpha 0} &= \int_{-\infty}^{\infty} \frac{d\omega}{2\pi} \, \Gamma_{\alpha} [1 - f_{\alpha} (\tilde{\xi}_{\alpha} + \omega)] F_{\alpha}^{>}(\omega) \, , \nn
\end{align}
where the tunneling-induced level broadening is $\Gamma_{\alpha} = 2\pi \rho_{\alpha} t_{\alpha}^{2}$ with $\rho_{\alpha}$ being
the constant density of states in lead $\alpha$ and $f_{\alpha}(x) = [e^{\beta_{el} (x-\mu_{\alpha})} +1]^{-1}$ is the Fermi distribution
function. Here, $\mu_{\alpha}$ is the chemical potential of lead $\alpha$ and $\beta_{el}$ denotes the inverse temperature of the
lead electrons. Note that we set $k_{B} = 1$.

In the steady state ($\dot{\rho}_{\alpha\beta} = 0$), the system given by Eqs.~(\ref{eqn:rateeqsD1})-(\ref{eqn:rateeqsD5}) can be
solved easily. The solution to the off-diagonal matrix elements in the steady state is given by
\begin{align}
	\rho_{LR} =	&- \frac{t_{D} M_{\trm{LR}}}{[ \tilde{\xi}_{L} - \tilde{\xi}_{R} ] - i \fw/2} \Big[ \rho_{RR} - \rho_{LL} \Big] \, , \nn \\
	\rho_{RL} =	&- \frac{t_{D} M_{\trm{RL}}}{[ \tilde{\xi}_{L} - \tilde{\xi}_{R} ] + i \fw/2 } \Big[ \rho_{RR} - \rho_{LL} \Big] \, \nn ,
\end{align}
which we use to write
\begin{align}
	0 &= - \Big[ W_{0 L} + W_{0 R} \Big] \rho_{00} + W_{L 0} \rho_{LL} + W_{R 0} \rho_{RR} \, , \label{eqn:rateeqs1} \\
	0 &= t_{D}^{2} \fv \left[\rho_{RR} - \rho_{LL}\right] - W_{L 0} \rho_{LL} + W_{0 L} \rho_{00} \, , \label{eqn:rateeqs2} \\
	0 &= t_{D}^{2} \fv \left[\rho_{LL} - \rho_{RR}\right] - W_{R 0} \rho_{RR} + W_{0 R} \rho_{00} \, , \label{eqn:rateeqs3}
\end{align}
where we defined
\begin{align} \label{eqn:o1}
	\fv = \frac{\fw M_{\trm{LR}} M_{\trm{RL}}}{ \fw^{2}/4 + (\tilde{\xi}_{L}-\tilde{\xi}_{R})^{2}} \, .
\end{align}

The stationary current can then obtained by
\begin{widetext}
\begin{align}
	I = -e \frac{ t_{D}^{2} \fv [ W_{0 L} W_{R 0} -W_{0 R} W_{L 0} ]}{t_{D}^{2} \fv [ 2 W_{0 L} + 2 W_{0 R} + W_{L 0} + W_{R 0}] + W_{0 R} W_{L 0} +  W_{0 L} W_{R 0} + W_{L 0} W_{R 0}} \, . \nn
\end{align}
\end{widetext}
The equation for the current nicely shows one major difference to the case of a single quantum dot coupled to a single bosonic mode,
viz. non-vanishing off-diagonal density-matrix elements. This allows coherent tunneling between the two dots. In Ref.~[\onlinecite{Zippilli2009}]
it was shown that in a double dot setup with a single bosonic mode such coherent tunneling can lead to cooling of the bosonic mode.

The influence of the phonons on the transport is due to $M_{\alpha \beta}$ and $F^{\lessgtr}_{\alpha}(\omega)$ which
are bosonic correlation functions.
The function $F^{<}_{\alpha}(t)$ is given by $F^{<}_{\alpha}(t) = \tr_{\trm{ph}}\left[ \rho_{\trm{ph}} X^{\pd}_{\alpha}(t) X^{\dag}_{\alpha} \right] = \ex{ X^{\pd}_{\alpha}(t) X^{\dag}_{\alpha}}$.
The Fourier transform is defined as $F^{<}_{\alpha}(\omega) = \int dt e^{i \omega t} F^{<}_{\alpha}(t)$. The greater
function can be obtained from the lesser function by the relation $F^{>}_{\alpha}(\omega) = F^{<}_{\alpha}(-\omega)$. Since, in the
derivation of the Born-Markov master equation, we assume equilibrated phonons, the expectation value of the bosonic
correlation functions is taken with respect to a thermal density matrix. In this case, the Fourier transform of $F^{<}_{\alpha}(t)$
can be calculated exactly \cite{Mahan2000}
\begin{align}
	F^{<}_{\alpha}(\omega) &= \sum_{n=-\infty}^{\infty} I_{n}\left[\frac{g_{\alpha}}{\sinh(\beta_{\trm{bos}}\Omega_{\alpha}/2)}\right] \exp\left[n \frac{\beta_{\trm{bos}} \Omega_{\alpha}}{2} \right] \nn \\
		\times& \exp\left[-g_{\alpha} \coth\left(\frac{\beta_{\trm{bos}} \Omega_{\alpha}}{2}\right) \right] 2 \pi \delta(\omega-n\Omega_{\alpha}) \nn \, ,
\end{align}
where, $I_{n}$ is the modified Bessel function of first kind, $g_{\alpha} = \Lambda_{\alpha}^{2} m_{\alpha} \Omega_{\alpha} /2 = \Lambda_{\alpha}^{2} / (2 \Lambda_{\alpha 0})$,
and $ \Lambda_{\alpha 0} = \sqrt{1/m_{\alpha} \Omega_{\alpha}}$. Here, $\beta_{\trm{bos}}$ is the inverse temperature of the phonon.
The correlation function $M_{\alpha \beta}$ is time independent and given by $M_{\alpha \beta} = \tr_{\trm{ph}}\left[ \rho_{\trm{ph}} X_{\alpha}^{\dag} X_{\beta}^{\pd} \right] = \ex{X_{\alpha}^{\dag} X_{\beta}^{\pd}}$.
For equilibrated phonons we have
\begin{align}
	M_{\alpha \beta} &= (1-e^{-\beta_{\rm{bos}} \Omega_{\alpha}}) e^{-g_{\alpha}/2} (1-e^{-\beta_{\rm{bos}} \Omega_{\beta}}) e^{-g_{\beta}/2} \nn \\
		\times& \sum_{n=0}^{\infty} e^{-\beta_{\rm{bos}} \Omega_{\alpha} n} L_{n}(g_{\alpha}) \sum_{m=0}^{\infty} e^{-\beta_{\rm{bos}} \Omega_{\beta} m} L_{m}(g_{\beta}) \nn \, ,
\end{align}
where $L_{n}$ are Laguerre polynomials.

\section{Current and differential conductance}
\label{sec:currentANDdIdV}

In the following, we study the current and the differential conductance through the double dot system. From now on we
assume, for simplicity, that both phonons have the same frequency $\Omega_{L} = \Omega_{R} = \Omega$. However, our
main results are not qualitatively affected by this assumption. A symmetric bias voltage is applied such that $\mu_{L} = V/2$
and $\mu_{R} = -V/2$. If not stated otherwise we choose $\beta_{el} = \beta_{\trm{bos}} = 10 \, \Omega$ for the electronic
and bosonic temperature, respectively. This corresponds to low temperatures for electrons in the leads as well as low
temperatures for the phonons. Put differently, $\beta_{\trm{bos}} = 10 \, \Omega$ means a low effective occupation number
of the phonon modes $n_{\trm{eff}} \approx 0$. As a consequence, the phonons can only absorb energy which is emitted
by the tunneling electron (and not emit energy to the electrons).
The coupling to the phonon modes opens additional transport channels. In particular, a tunneling electron can now emit
a phonon during the tunnel process (the absorption process is suppressed because of $n_{\trm{eff}} \approx 0$). This
emission process leads to additional steps in the $I(V)$ curve or equivalently to additional resonances in the differential
conductance $dI/dV$.

\subsection{Results}

\begin{figure}[t]
	\centering
	\includegraphics[width=0.99\columnwidth]{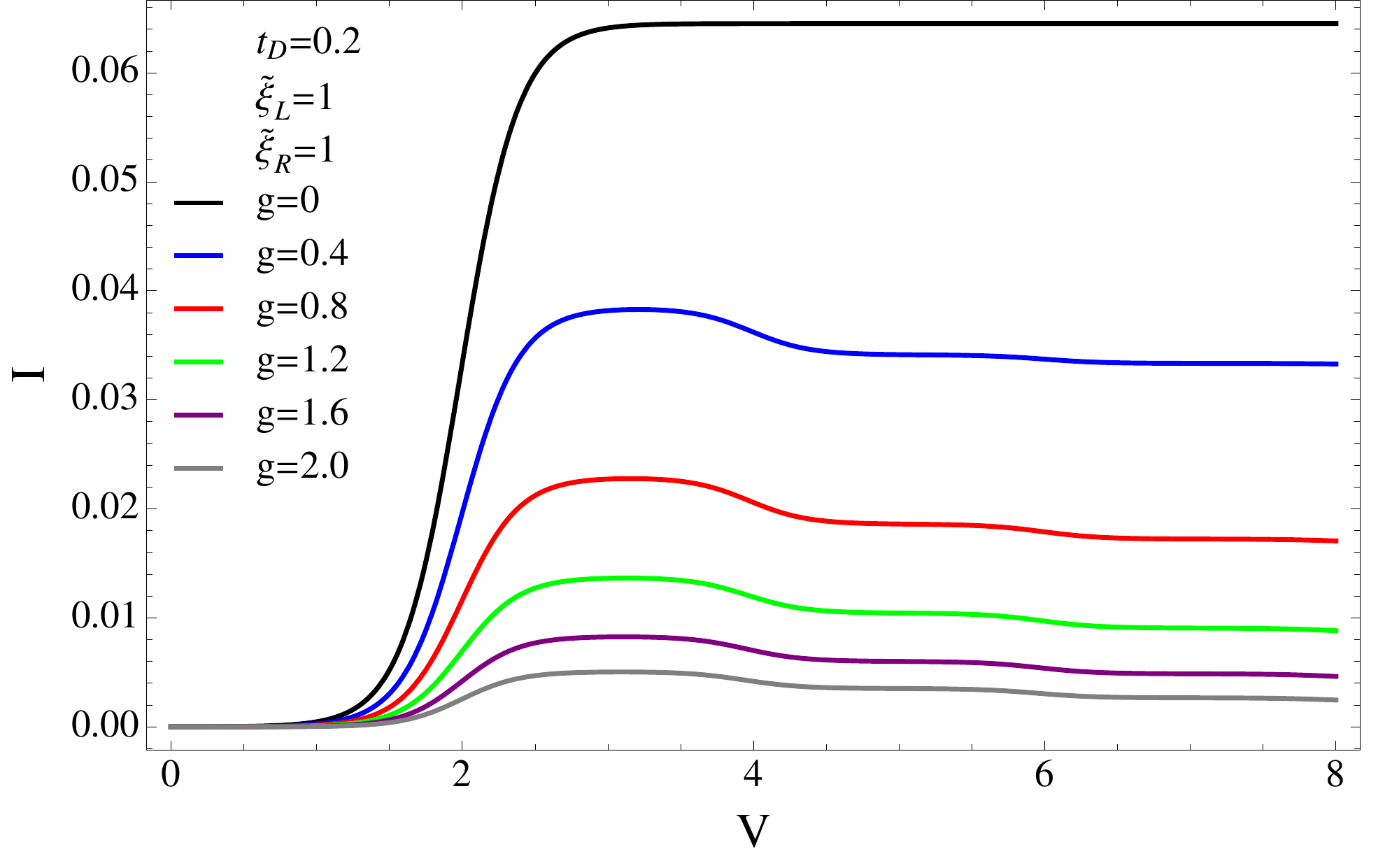}
	\caption{\label{fig:IV1} (Color online)
	Current $I$ through the double dot system as a function of bias voltage $V$ for different values of electron-phonon
	coupling $g$ and symmetric level energies $\tilde{\xi}_{L} = \tilde{\xi}_{R}$. The current decreases with increasing
	electron-phonon coupling. For a fixed (nonzero) value of the electron-phonon coupling, the current also decreases
	(in some regions) with increasing bias voltage. Parameters are given in the legend. Here, and in the following
	figures, energies are in units of $\Omega$. Hence, the bias voltage $V$ is shown in units of $\Omega/e$ and the current $I$ in units of $(e/h)\Omega$.
	}
\end{figure}

\begin{figure}[t]
	\centering
	\includegraphics[width=0.99\columnwidth]{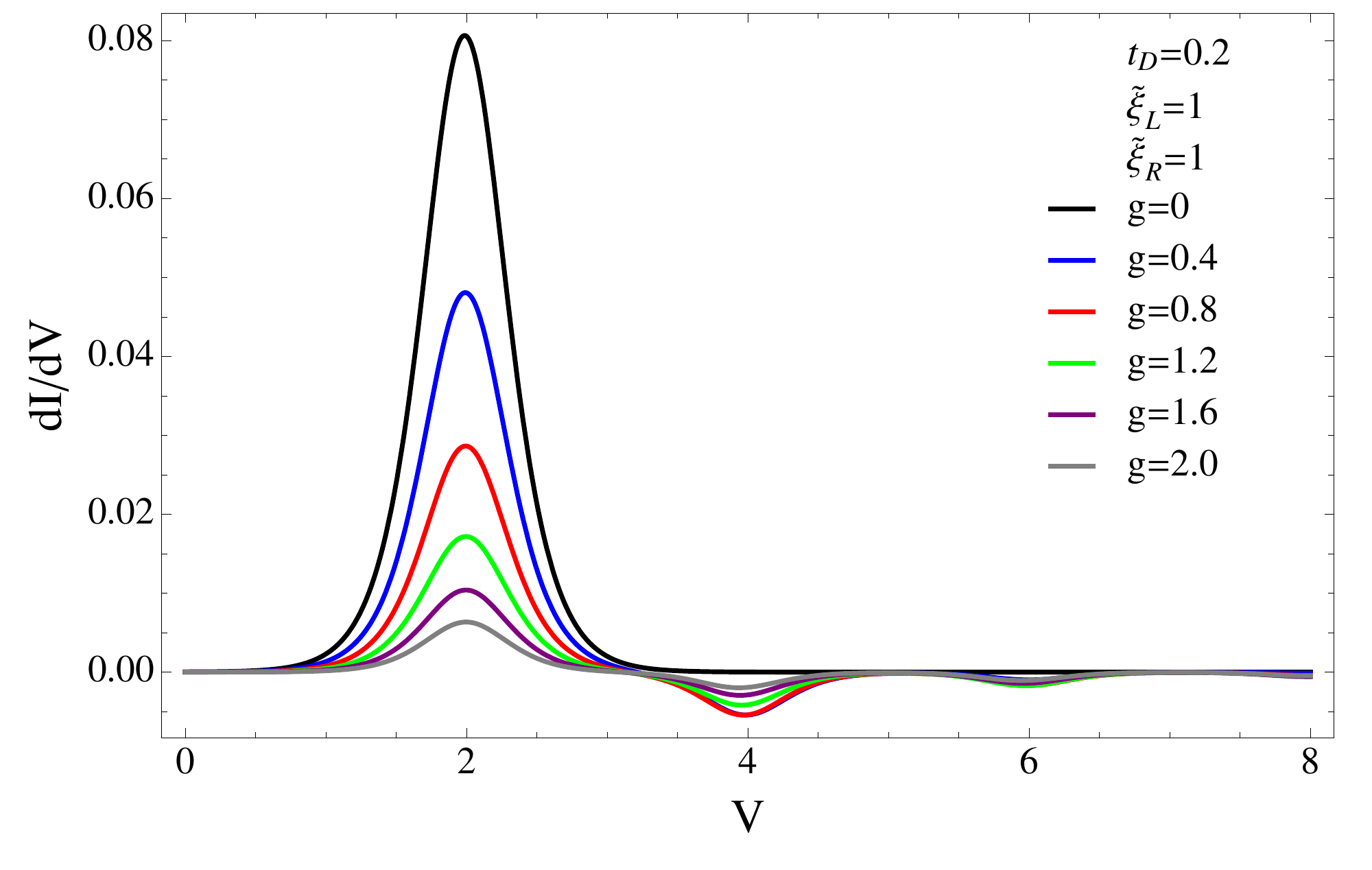}
	\caption{\label{fig:dIdV1} (Color online)
	Differential conductance $dI/dV$ for the same parameters as in Fig.~\ref{fig:IV1}.
	We clearly see the negative differential conductance.
	}
\end{figure}

\begin{figure}[t]
	\centering
	\includegraphics[width=0.99\columnwidth]{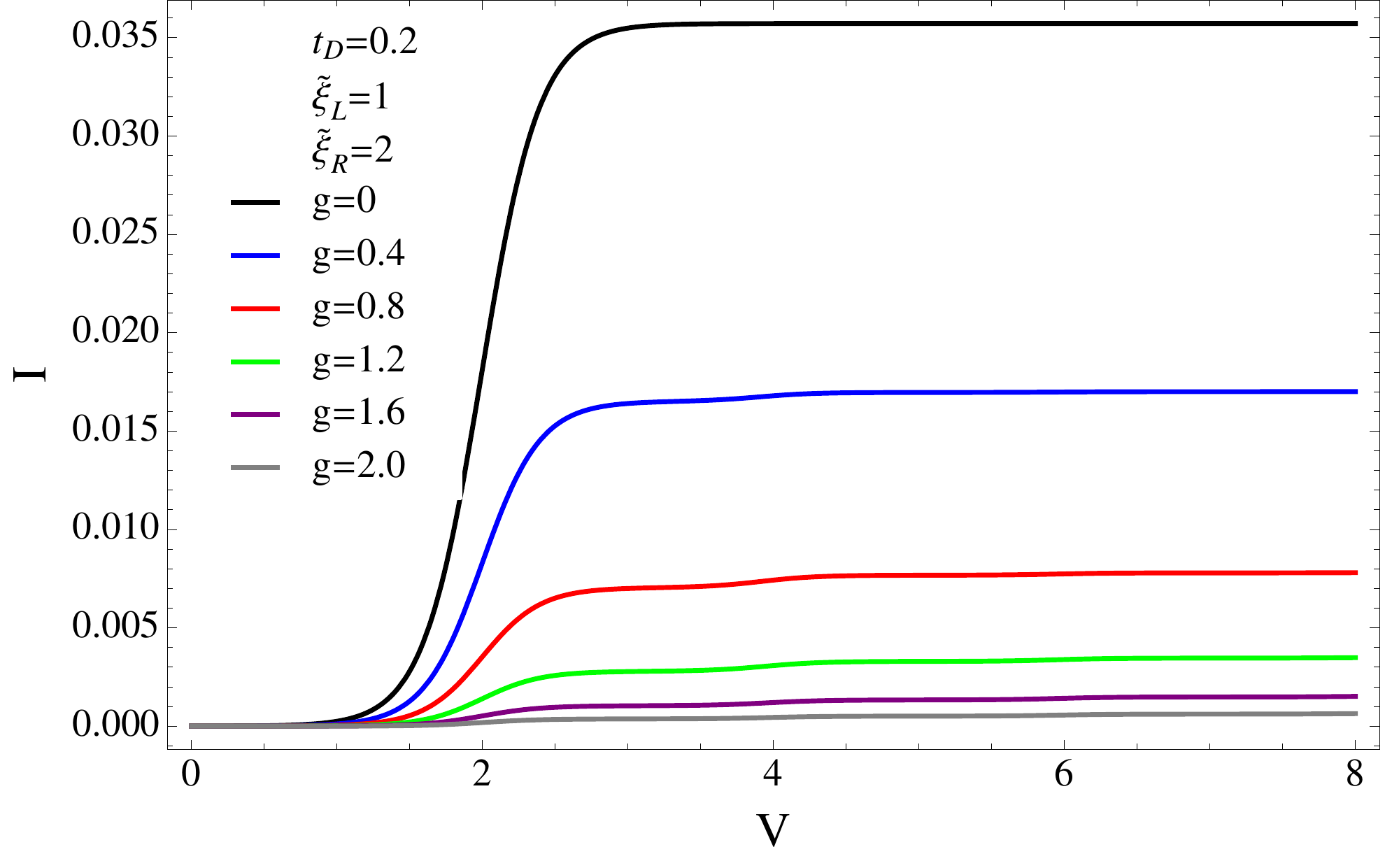}
	\caption{\label{fig:IV2} (Color online)
	Current $I$ through the double dot system as a function of bias voltage $V$ for different electron-phonon
	couplings $g$ and an asymmetric choice of level energies $\tilde{\xi}_{L} \neq \tilde{\xi}_{R}$. In comparison
	to Fig.~\ref{fig:IV1}, the current for a fixed electron-phonon coupling is now increasing with the bias. All
	parameters are given in the legend.
	}
\end{figure}

\begin{figure}[t]
	\centering
	\includegraphics[width=0.99\columnwidth]{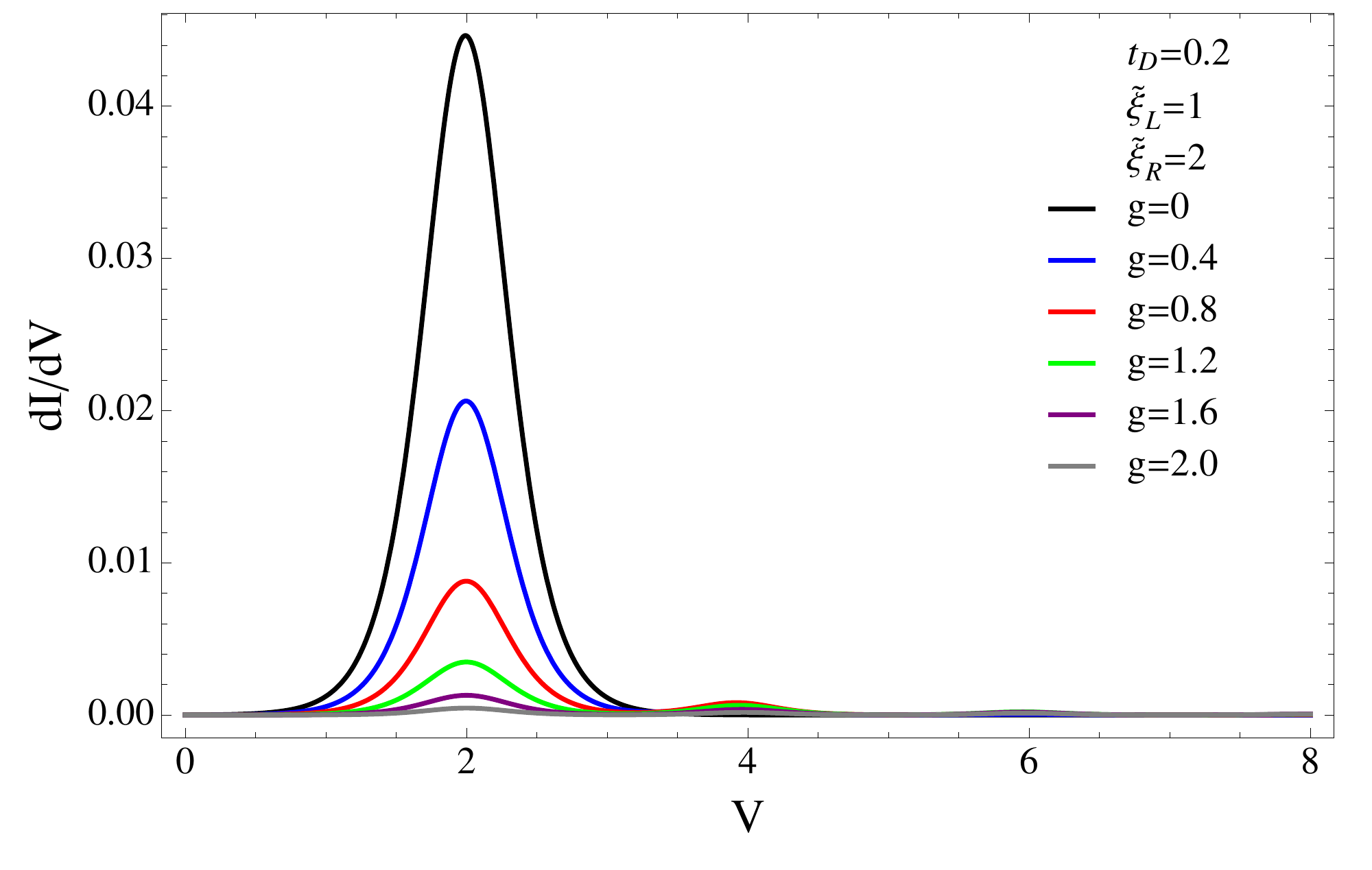}
	\caption{\label{fig:dIdV2} (Color online)
	Here, we show the differential conductance for the same parameters as in Fig.~\ref{fig:IV2}.
	We clearly see the absence of a negative differential conductance.
	}
\end{figure}

In Figs.~\ref{fig:IV1}-\ref{fig:dIdV2}, we present our results on the current through and the differential conductance of
the double quantum dot system. In the following, we chose, again for simplicity, a symmetric electron-phonon coupling $g_{L}=g_{R}=g$.

In Fig.~\ref{fig:IV1}, we show the current through the double dot system as a function of bias voltage $V$ for different
values of the electron-phonon coupling $g$ and for aligned left and right electronic levels ($\tilde{\xi}_{L} = \tilde{\xi}_{R}$).
First, we see that for fixed bias voltage the current decreases towards stronger electron-phonon coupling. Furthermore,
for a fixed electron-phonon coupling, the current beyond a certain bias voltage also decreases upon increasing the bias
voltage. This leads to a negative differential conductance, which can be seen in more detail in Fig.~\ref{fig:dIdV1}. The
steps in the current (peaks in the differential conductance) appear whenever an electron can emit a phonon while
tunneling. The current obeys the symmetry $I(V) = -I(-V)$ due to our symmetric choice of parameters.

Figure~\ref{fig:IV2} shows $I(V)$ for different electron-phonon couplings $g$ in the case of asymmetric level energies
$\tilde{\xi}_{L} - \tilde{\xi}_{R} \approx \Omega$. This asymmetry can, for instance, be induced by tuning the dot level energies
$\tilde{\xi}_{\alpha}$ with a gate voltage. Due to the asymmetry in the setup, the current is then no longer an antisymmetric
function of voltage, $I(V) \neq -I(-V)$. As before, a stronger electron-phonon coupling leads to a decrease of the
current. However, the current for fixed electron-phonon coupling now always increases with the bias voltage. Therefore,
introducing an asymmetry in the setup causes the negative differential conductance to disappear, see Figs.~\ref{fig:IV2}
and~\ref{fig:dIdV2}. For the differential conductance to become positive, the introduced asymmetry has to be of the order
$\tilde{\xi}_{L} - \tilde{\xi}_{R} \approx \Omega$, see the next section for an explanation why.

Figures~\ref{fig:IV1}-\ref{fig:dIdV2} are the first main result of our article, showing that electron-phonon coupling in a
double quantum dot can lead to a negative differential conductance, and that this effect can be influenced by adjusting
the level energies.
A different way to remove the negative differential conductance is to increase the inter-dot tunneling $t_{D}$ which leads
to an increased tunneling rate between the dots. We discuss the nature and origin of the negative differential conductance
in the next section.

\subsection{Origin of the negative differential conductance}

In the absence of electron-phonon coupling, the tunneling-induced width of the dot levels allows for transport through the
double quantum dot even in an off-resonant situation. The tunneling rate between the left and right dot can be associated
with $\fv$. According to Eq.~(\ref{eqn:o1}),
\begin{align}
	\fv(g=0) = \frac{\Gamma_{L} + \Gamma_{R}}{ (\Gamma_{L} + \Gamma_{R})^{2}/4 + (\tilde{\xi}_{L}-\tilde{\xi}_{R})^{2}} \, . \nn
\end{align}
This can be interpreted as the density of states of the left dot at the energy of the right one. $\fv(g=0)$ depends only on
$\Gamma_{\alpha}$ and the energy difference of the levels. If the levels are aligned, $\tilde{\xi}_{L}=\tilde{\xi}_{R}$, $\fv$
reaches its maximum and so does the current. On the other hand, $\fv$ and the current, both decrease if
the energy difference of the levels $\tilde{\xi}_{L}-\tilde{\xi}_{R}$ is nonzero.
Therefore, without phonons the differential conductance (the peak height and width) is predominantly described by the
tunneling-induced level broadening $\Gamma_{\alpha}$ and the level energies.

\begin{figure}[t]
	\centering
	\includegraphics[width=0.99\columnwidth]{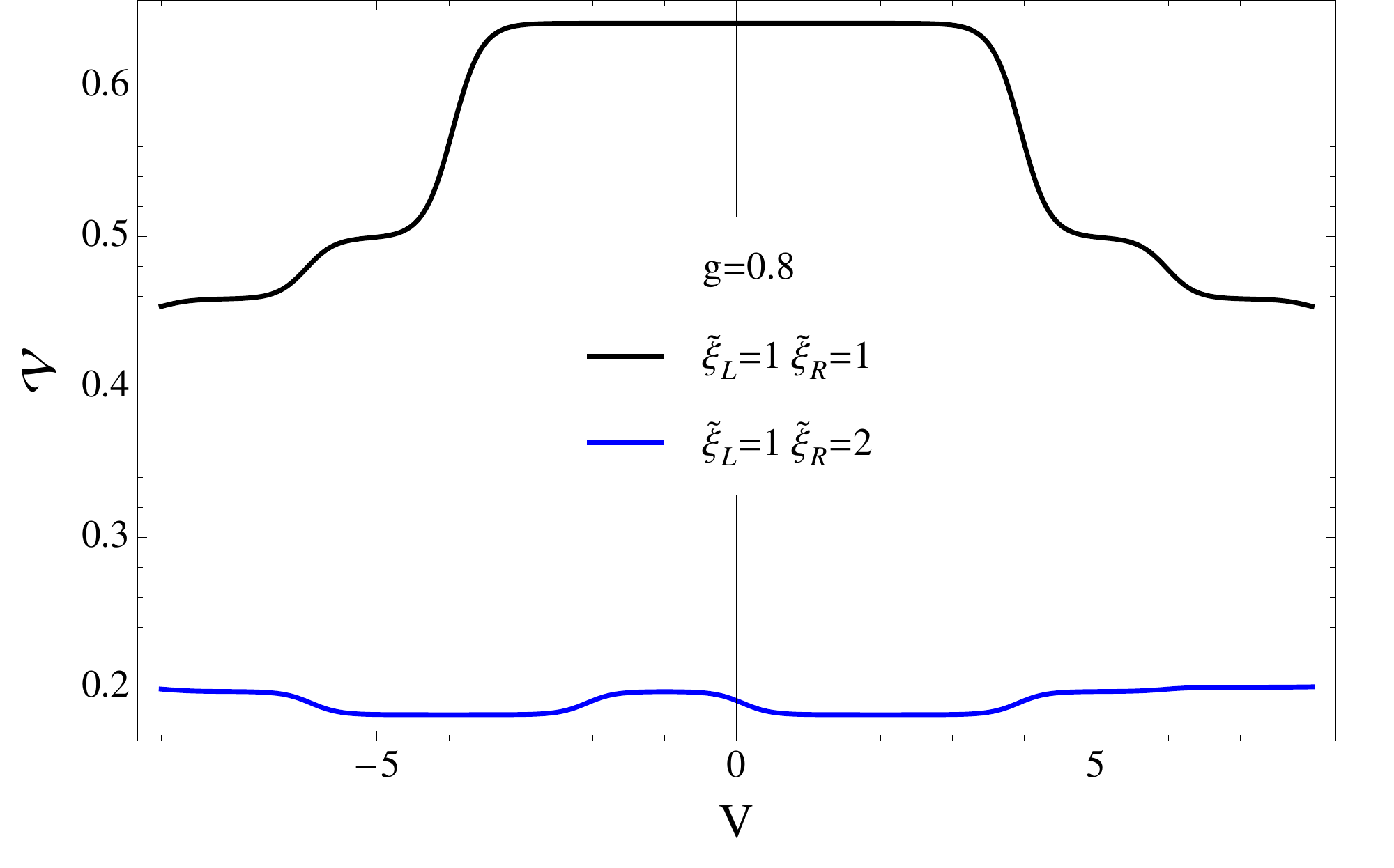}
	\caption{\label{fig:fV} (Color online)
	The function $\fv$ as a function of $V$ for $\tilde{\xi}_{L}=\tilde{\xi}_{R}=1$ (black) and $\tilde{\xi}_{L}=1$,
	$\tilde{\xi}_{R}=2$ (blue). For $\tilde{\xi}_{L}=\tilde{\xi}_{R}=1$, $\fv$ decreases at every phonon sideband in
	contrast to the case $\tilde{\xi}_{L}=1$, $\tilde{\xi}_{R}=2$. In this case at every phonon sideband (the first
	one being at $V \approx 4$) $\fv$ increases. (Note the blue curve is shifted to the left due to the asymmetry
	in the setup.)
	}
\end{figure}

In the case of nonzero electron-phonon coupling, the situation is very different.
Most importantly, due to the presence of phonons, $\fv$ depends on the bias voltage. We also know from
Figs.~\ref{fig:dIdV1} and~\ref{fig:dIdV2} that we have to distinguish the cases $\tilde{\xi}_{L} = \tilde{\xi}_{R}$ and $\tilde{\xi}_{L} \neq \tilde{\xi}_{R}$.
First, for aligned levels $\tilde{\xi}_{L} = \tilde{\xi}_{R}$ we obtain
\begin{align}
	\fv(\tilde{\xi}_{L} = \tilde{\xi}_{R}) = 4 \, \frac{M_{\trm{LR}} M_{\trm{RL}}}{ \fw } \, . \nn
\end{align}
The bias voltage only enters in $\fw$, which increases whenever the bias voltage reaches a phonon sideband. Therefore,
$\fv$ decreases at these thresholds, see Fig.~\ref{fig:fV}. If we interpret $\fv$ again as a density of states, this decrease
indicates that due to the phonons fewer states are available for transport. Second, in the case of a finite energy difference
of the levels of order $\tilde{\xi}_{L} - \tilde{\xi}_{R} \approx \Omega$, the rate becomes approximately
\begin{align}
	\fv(\tilde{\xi}_{L} - \tilde{\xi}_{R} \approx \Omega) \sim \fw M_{\trm{LR}} M_{\trm{RL}} \, . \nn
\end{align}
In this case, the rate $\fv$ increases with the bias voltage at each phonon sideband. In Fig.~\ref{fig:fV}, we show $\fv$ as a
function of the bias voltage for the two cases discussed above. To summarize, this explains the
occurrence of negative differential conductance at large electron-phonon coupling, and why it disappears when the
inter-dot tunneling is increased.

The negative differential conductance can be explained physically as follows. If the bias voltage exceeds the phonon frequency, tunnel processes
become possible in which the electron emits a (real) phonon when entering, say, the left dot. As a consequence, its energy may be 
insufficient to tunnel to the right dot, so transport is blocked. Ultimately, the electron will escape again from the
left dot, either by reabsorbing the phonon or by co-tunneling directly to the right reservoir. This short blockade of transport leads to a decrease of the total current once the bias
voltage exceeds the phonon frequency, and hence to a negative differential conductance.

There is a stark contrast between the double dot setup with phonons and a single-level quantum dot that couples to one
phonon mode. When phonons are involved in the transport through a single quantum dot the so-called Franck-Condon
blockade \cite{Koch2005} arises. Then, in the sequential tunneling limit, the differential conductance is positive and the
current through the single quantum dot is suppressed for low bias voltages when increasing the electron-phonon coupling.
Negative differential conductance due to phonons in a single single-level quantum dot is only possible due to higher order co-tunneling
processes~\cite{Koch2006} or asymmetric coupling of the dot to the leads~\cite{Zazunov2006}.

\subsection{Occupation probabilities}

\begin{figure}[t]
	\centering
	\includegraphics[width=0.95\columnwidth]{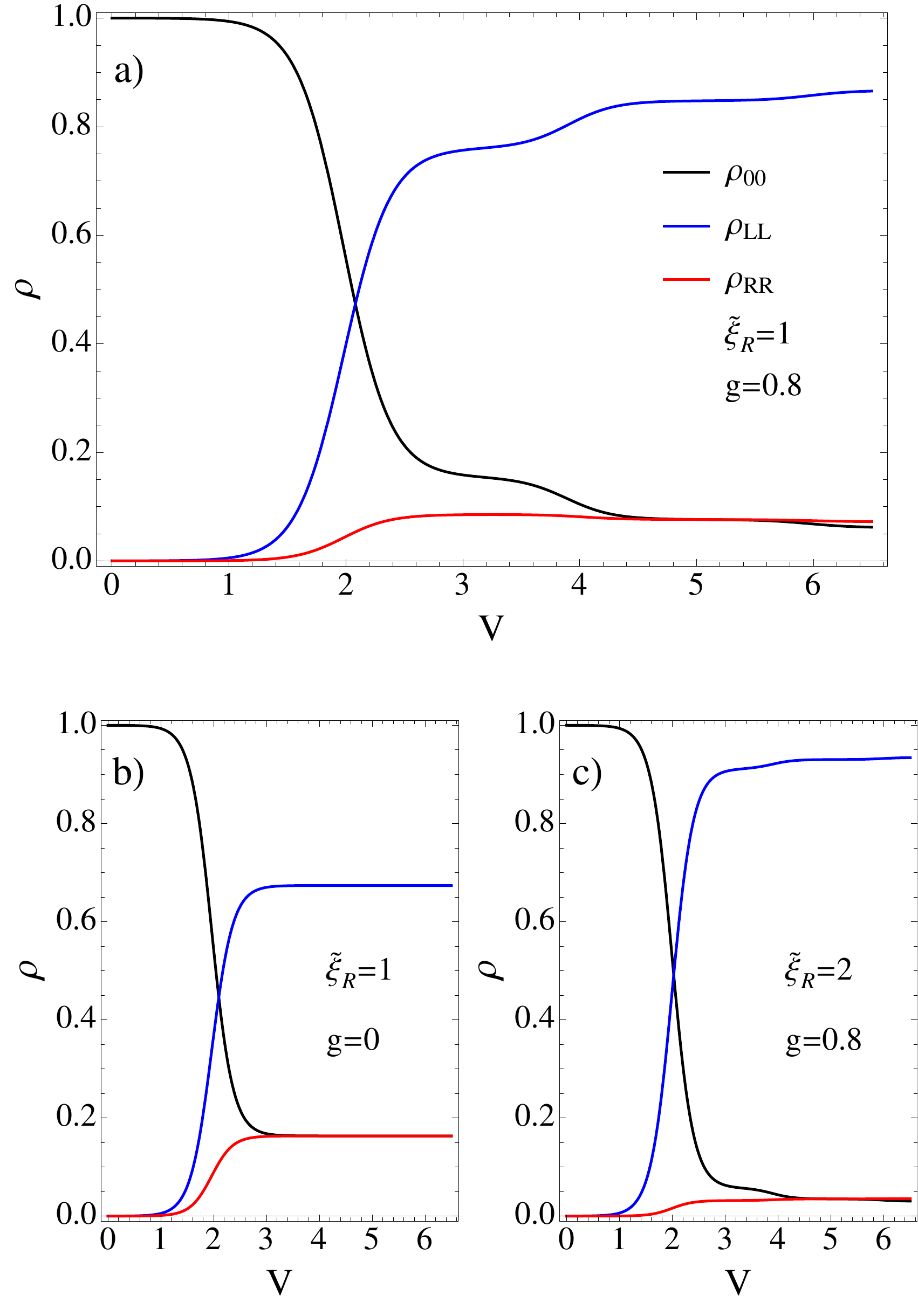}
	\caption{\label{fig:probs} (Color online)
	Diagonal elements of the density matrix $\rho_{00}$ (black), $\rho_{LL}$ (blue), and $\rho_{RR}$ (red).
	In b) $g=0$ and level energies are chosen symmetricly. In a) and c) $g=0.8$ and level energies are chosen
	symmetric and asymmetric, respectively. Here, $t_{D} = 0.4$ and $\tilde{\xi}_{L}=1$.
	}
\end{figure}

An investigation of the occupation probabilities of the dot states further strengthens the explanation for the occurrence of
a negative differential conductance. Figure~\ref{fig:probs} shows the occupation probabilities of the dot states, i.e., the
diagonal elements of the dot density matrix.

In Fig.~\ref{fig:probs}b), we see that without electron-phonon coupling and $\tilde{\xi}_{L} = \tilde{\xi}_{R}$, the probability
for having zero electrons in the double dot ($\rho_{00}$) decreases when the bias voltage $V$ is increased.
Simultaneously, the probabilities $\rho_{LL}$ and $\rho_{RR}$ both increase. As a consequence the current through
the system increases until it saturates.

For nonzero electron-phonon coupling ($g=0.8$) and $\tilde{\xi}_{L} = \tilde{\xi}_{R}$, on the other hand, we recognize from
Fig.~\ref{fig:probs}a) that at the first phonon sideband, the occupation probability of the left dot increases but the
occupation probability of the right dot decreases. This behavior suggests that the inter-dot transport from the left to the right
dot becomes suppressed at this bias voltage threshold. Thus, the current decreases when the bias voltage is increased
beyond the threshold voltage which is the onset of a negative differential conductance.

In Fig.~\ref{fig:probs}c), $\tilde{\xi}_{L} \neq \tilde{\xi}_{R}$ and the other parameters are the same as in Fig.~\ref{fig:probs}a).
At the first phonon sideband the occupation probability of the left dot increases (as before) and now the occupation
probability of the right dot also increases. This behavior is qualitative similar to the one depicted in Fig.~\ref{fig:probs}b)
and therefore the differential conductance is purely positive.\\

\section{Summary}
\label{sec:summary}

To summarize, we have investigated transport properties, namely the current and the differential conductance, in a
double quantum dot setup in which a phonon mode is coupled to each quantum dot.
We have shown that the electron-phonon coupling gives rise to a negative differential conductance under certain conditions. Furthermore, we have
argued that the electron-phonon coupling leads to an inter-dot tunneling rate that depends on the bias voltage and on
the energy difference between the dots, which we identified as the origin of the occurrence of negative differential conductance.
The very generic model we used can readily be probed in nano-electromechanical systems. Experiments employing
suspended carbon nanotubes incorporate both, single localized levels and phonon modes. In addition to that, strong
electron-phonon coupling, high $Q$-factors, and high resonance frequencies make carbon nanotubes perfect candidate
devices to study the occurrence of negative differential conductance in double-quantum dot systems with electron-phonon
coupling.

\section*{Acknowledgements}

We would like to thank Christoph Bruder, Shahal Ilani, and Christoph Stampfer for stimulating discussions. Financial
support by the Swiss SNF, the NCCR QSIT, and the German DFG is gratefully acknowledged.



\end{document}